\documentclass[aps,pre,twocolumn,groupedaddress]{revtex4-1}

\usepackage{amsmath, amsfonts, amssymb}
\usepackage{mathtools} 
\usepackage[pdftex,colorlinks,bookmarks = true]{hyperref} 
\usepackage{siunitx} 
\usepackage{csquotes} 
\usepackage{braket} 
\newcommand{\avg}[1]{\left< #1 \right>} 
\newcommand{\inv}{^{-1}} 

\begin{document}

\title{Multihistogram Reweighting for Nonequilibrium Markov Processes Using Sequential Importance Sampling Methods}
\author{Troels Arnfred Bojesen}
\email[]{troels.bojesen@ntnu.no}
\affiliation{Department of Physics, Norwegian University of Science and Technology, N-7491 Trondheim, Norway}
\date{\today}

\begin{abstract}
We present a multihistogram reweighting technique for nonequilibrium Markov Chains with discrete energies. The method generalizes the single histogram method of Yin et al. [Phys. Rev. E \textbf{72}, 036122 (2005)], making it possible to calculate the time evolution of observables at a posteriori chosen couplings based on a set of simulations performed at other couplings. In the same way as multihistogram reweighting in an equilibrium setting improves the practical reweighting range as well as use of available data compared to single histogram reweighting, the method generalizes the multihistogram advantages to nonequilibrium simulations. We demonstrate the procedure for the Ising model with Metropolis dynamics, but stress that the method is generally applicable to a range of models and Monte Carlo update schemes.
\end{abstract}


\maketitle

\section{Introduction}
In the last two decades the use of short time critical dynamics (STCD) as a way of investigating the critical properties of models in statistical mechanics has emerged as an interesting alternative to equilibrium Monte Carlo (EMC) simulations. The reason is twofold: First, STCD simulations may, given sufficiently large system sizes, effectively avoid finite size effects, since the correlation length is still small in the short time regime. Second, and even more important, critical slowing down is avoided due to the system being far from equilibrium. See for instance Refs. \onlinecite{Ozeki_Ito_2007,Albato_et_al_2011} for some recent reviews on the STCD method.

The EMC simulations have, on the other hand, long had the advantage of the powerful histogram reweighting methods of Ferrenberg and Swendsen, meaning that knowledge of the exact couplings of interest are not needed a priori the simulations: Single histogram reweighting~\cite{PhysRevLett.61.2635} makes it, in principle, possible to extract observables at arbitrary couplings from a simulation performed at a fixed coupling, although the practical range is limited to a small neighborhood of the simulation coupling. The practical reweighting range as well as the statistical quality of the observables can be improved by using multihistogram reweighting~\cite{PhysRevLett.63.1195}, where data from an arbitrary number of simulations at different couplings are combined in an optimal way.

A first attempt to introduce reweighting techniques to STCD simulations was done by Lee and Okabe~\cite{PhysRevE.71.015102}. They proposed what essentially is (Monte Carlo) time dependent single histogram reweighting, where the reweighting is done simultaneously with the simulation. This is a major drawback of the method, as all the desired couplings, simulation as well as reweighted, must be known a priori. In addition, obtaining an observable through reweighting is almost as computationally heavy as performing the simulation itself, and since the procedure scales linearly with the number of reweighting couplings, the practical usefulness of the method is limited.

Yin et al. proposed a novel method that improves the single histogram reweighting for STCD, but the method is restricted to models with a discrete energy spectrum~\cite{PhysRevE.72.036122}: By keeping track of a time dependent histogram of the energy changes at each step in the Markov chain, and sampling these each time an observable is sampled, it is possible to efficiently reweight observables to a posteriori chosen couplings. This is a significant improvement, but the reweighting is still limited to the single histogram regime where only data from simulations at one coupling strength is used. In a real world situation one would typically perform several simulations at different couplings when searching for the critical one, thus obtaining a lot of data that contains potentially useful information. It would therefore be advantageous to develop a multihistogram reweighting technique for nonequilibrium Markov processes.

In this paper we propose such a technique for nonequilibrium reweighting of models with \emph{discrete energy spectra} by extending the method of Yin et al. to time dependent multihistogram reweighting using the method of Ferrenberg and Swendsen. After the general derivation we test the method on a Metropolis dynamics STCD simulation on the 2D Ising model and compare it with the \enquote{raw} (non-reweighted) as well as single histogram reweighted results.

Note that although these methods are developed with STCD simulations in mind, they are generally applicable to all nonequilibrium Markov processes (and hence also equilibrium processes).

\section{Multihistogram Reweighting}

\subsection{General Formulation}
Let $x_t = (\sigma_{1},\sigma_{2},\ldots,\sigma_{t})$ denote a Markov chain of field configurations $\sigma$ after $t$ steps (often called \enquote{time} from now on), and let $\Sigma_t$ be the dynamical phase space, i.e. the set of all \emph{possible} Markov chains of $t$ steps, starting from an initial set of states $\Sigma_0$. Given a (usually small) set of coupling parameters $\beta$, we may associate a probability weight to each Markov chain:
\begin{equation}
 w: \beta \times \Sigma_t \to \mathbb{R}.
 \label{eq:markov_chain_weight}
\end{equation}
At each time step we have a probability distribution over all possible \emph{reachable} states, which we may capture in the \emph{dynamical} partition function
\begin{equation}
 Z_t(\beta) \equiv \sum_{x_t} w(\beta,x_t),
 \label{eq:dynamical_partition_function}
\end{equation}
where $x_t$ under the summation sign is a shorthand notation for $x_t \in \Sigma_t$. In the same way we define the dynamical average of an observable $O$ as
\begin{equation}
 \avg{O}_t(\beta) \equiv Z_t(\beta)\inv \sum_{x_t} w(\beta,x_t)O(\sigma_t).
 \label{eq:dynamical_average}
\end{equation}
Thus, at a fixed time $t$ there is no difference between the dynamical formalism and the equilibrium formalism, except that the fundamental \enquote{states} in the dynamical case are entire Markov chains instead of single field configurations. With this in mind, methods from equilibrium simulations may be carried right over to dynamic simulations.

For the sake of simplicity, we will from now on restrict ourselves to the special case when the set of coupling parameters consists of only one coupling parameter (also named $\beta$). Generalization to multiple parameters is straightforward.

In the equilibrium formalism, Ferrenberg--Swendsen multihistogram reweighting goes as follows~\cite{PhysRevLett.63.1195}: Let $H(\sigma) = H_0(\sigma) + \beta S(\sigma)$ be a general Hamiltonian of a field $\sigma$ and the partition function $Z(\beta) = \sum_\sigma \exp(H(\sigma))$. $S$ is an operator (energy, magnetization, etc.) on the field. Given $Q$ EMC simulations performed at couplings $\set{\beta_1,\ldots,\beta_q,\ldots,\beta_Q}$, each with $N_q$ samples with autocorrelation time $\tau_q$, the best approximation for the probability weight of $S$ at a (a posteriori chosen) coupling $\beta^*$ is given by
\begin{equation}
 P(\beta^*,S) = \frac{\sum_{q} (1+2\tau_q)\inv M_q(S) \exp(\beta^* S)}{\sum_{q} N_q (1+2\tau_q)\inv \exp(\beta_q S) \tilde{Z}(\beta_q)\inv},
 \label{eq:classic_MHRW}
\end{equation}
where $M_q(S)$ is the histogram of the $S$-measurements of the $q$'th series, and
\begin{equation}
 \tilde{Z}(\beta) = \sum_S P(\beta,S).
 \label{eq:classic_Z}
\end{equation}
$P$ is found by solving Eqs. \eqref{eq:classic_MHRW} and \eqref{eq:classic_Z} iteratively. The error minimized average of an observable (operator on $S$) at $\beta^*$ is then given by
\begin{equation}
 \avg{O}(\beta^*) = \tilde{Z}(\beta^*)\inv \sum_S O(S) P(\beta^*,S).
\end{equation}

We now turn to the dynamic generalization of this method. Let $X_{q,t} \equiv \set{x_{q,t}}$ be a set of $N_q$ Markov chains obtained from independent simulations at the same coupling $\beta_q$, and let $X_t \equiv \set{X_{q,t}|q \in \{1,\ldots, Q\}}$ be the collection of all Markov chains from $Q$ such sets, each with their own associated coupling $\beta_q \in \set{\beta_1,\ldots,\beta_Q }$. Replacing $\exp(\beta S) \to w(\beta,x_t)$, $S \to x_t$ (i.e. $x_t$ may be seen as the identity operator), $P(\beta,S) \to W(\beta,x_t)$, and $\tilde{Z}(\beta) \to \tilde{Z}_t(\beta) = \sum_{x_t \in X_t} W(\beta,x_t)$ in Eq. \eqref{eq:classic_MHRW}, we get
\begin{equation}
 W(\beta^*,x_t) = \frac{\sum_q M_q(x_{t})w(\beta^*,x_t)}{\sum_{q} N_q w(\beta_q, x_t) \tilde{Z}_t(\beta_q)\inv}.
 \label{eq:dynamical_MHW_0}
\end{equation}
$\tau_q = 0$ since the Markov chains are independent. $M_q(x_{t})$ is the histogram of Markov chains on the form $x_{t}$ for $X_{q,t}$, i.e. $M_q(x_{t}) = \sum_{y\in X_{q,t}} \delta_{y,x_t}$. Now, if we treat all the obtained Markov chains as unique (even in the improbable case of two being identical), there will be \emph{only one} $y \in X_t$ fulfilling $y = x_t$, namely $x_t$ itself. Hence we must have that $\sum_q M_q(x_{t}) = 1$, and Eq. \eqref{eq:dynamical_MHW_0} simplifies to
\begin{equation}
 W(\beta^*,x_t) = \frac{w(\beta^*,x_t)}{\sum_{q} N_q w(\beta_q, x_t) \tilde{Z}_t(\beta_q)\inv}.
 \label{eq:dynamical_MHW}
\end{equation}
The error minimizing dynamical average is then
\begin{equation}
 \avg{O}_t(\beta^*) = \tilde{Z}_t(\beta^*)\inv \sum_{x_t} O(\sigma_t) W(\beta^*,x_t),
 \label{eq:dynamical_average_MHRW}
\end{equation}
which is what we are interested in. To get any further with a practical use of Eqs. \eqref{eq:dynamical_MHW} and \eqref{eq:dynamical_average_MHRW} we need to determine the Markov chain weights $\set{w(\beta,x_t)}$.

\subsection{The Markov Chain Weight}
The weight of a Markov chain [Eq. \eqref{eq:markov_chain_weight}] is per definition given by
\begin{equation}
 w(\beta,x_t) = \omega(\sigma_0)\prod_t \omega(\beta,\sigma_t | \sigma_{t-1}),
 \label{eq:markov_chain_weight_general}
\end{equation}
where $\omega(\sigma_0)$ is the weight of the initial state and $\omega(\beta,\sigma_t | \sigma_{t-1})$ is the weight associated with a Markov step from $\sigma_{t-1}$ to $\sigma_t$ at coupling $\beta$. Since each Markov chain in an STCD simulation may consist of a very large number of steps (e.g. $10^{10}$ for a $10^{4}$ sweep simulation on a $1000 \times 1000$ lattice), it is practically impossible to directly store the weight information of all steps for later reweighting; in the general case, where the set of possible step weights is large (possibly infinite), one has no choice but to reweight \enquote{on the fly}, as Lee and Okabe proposed. 

If, however, we restrict ourselves to models where the set of possible step weights is discrete, it \emph{is} possible to find the the Markov chain weight at an a posteriori chosen coupling $\beta$. The trick is to write Eq. \eqref{eq:markov_chain_weight_general} as~\cite{PhysRevE.72.036122}
\begin{equation}
 w(\beta,x_t) = \omega(\sigma_0)\prod_{\alpha}\omega(\beta,\alpha)^{r_{\alpha,t}},
 \label{eq:Markov_chain_weight_discrete}
\end{equation}
$\omega(\beta,\alpha)$ being the weight of a step of type $\alpha$, and $r_{\alpha,t} \in \mathbb{N}_0$ the number of such steps in the Markov chain $x_t$. $\omega(\beta,\alpha)$ can be calculated for an arbitrary $\beta$, given the knowledge of $\alpha$ and the dynamics used in the simulation. Thus, to know $w(\beta,x_t)$ we only need to know the initial state $\sigma_0$ and the histogram of $r_{\alpha,t}$-values. 

As an example, consider an STCD simulation of a spin model using canonical Metropolis dynamics at some coupling $\beta_q$. A trial update of a spin will result in an energy change $\Delta E_{\alpha}$, which we assume to be member of a discrete set of all possible (one step) energy changes $\set{\Delta E}$. There are three possible outcomes of a trial update in the Metropolis dynamics: (See Ref. \onlinecite{PhysRevE.71.015102} for mathematical details.)
\begin{itemize}
 \item $\Delta E_{\alpha} \leq 0$, in which case the trial is always accepted. Then $\omega(\beta,\alpha) = 1$.
 \item $\Delta E_{\alpha} > 0$ and the update is accepted. Then $\omega(\beta,\alpha) = \exp(-\beta \Delta E_\alpha)$.
 \item $\Delta E_{\alpha} > 0$ and the update is rejected. Then $\omega(\beta,\alpha) = 1 - \exp(-\beta \Delta E_\alpha)$.
\end{itemize}
The explicit expression for the Markov chain weight, Eq.~\eqref{eq:Markov_chain_weight_discrete}, becomes
\begin{align}
 w(\beta,x_t) &= \omega(\sigma_0) \prod_{\substack{\alpha \\ \Delta E_\alpha > 0 \\ \text{accept} }} \left[\exp(-\beta \Delta E_\alpha)\right]^{r_{\alpha,t}} \nonumber \\ 
 {}&\qquad \times \prod_{\substack{\alpha \\ \Delta E_\alpha > 0 \\ \text{reject} }} \left[1 - \exp(-\beta \Delta E_\alpha)\right]^{r_{\alpha,t}} \nonumber \\
 &= \omega(\sigma_0) \exp(-\beta \Delta_t E^{\text{a}}) \nonumber \\
 {}&\qquad \times \prod_{\substack{\alpha \\ \Delta E_\alpha > 0 \\ \text{reject} }} \left[1 - \exp(-\beta \Delta E_\alpha)\right]^{r_{\alpha,t}},
\end{align}
where
\begin{equation}
 \Delta_t E^{\text{a}} \equiv \sum_{\substack{\alpha \\ \Delta E_\alpha > 0 \\ \text{accept} }} r_{\alpha,t} \Delta E_\alpha = \sum_{\substack{t \\ \Delta_t E > 0 \\ \text{accept} }} \Delta_t E
\end{equation}
is the running sum of all positive acceptance energy changes. So, to be able to calculate the Markov chain Metropolis dynamics weight at an arbitrary $\beta$, and thus to be able to multihistogram reweight an observable average using Eq. \eqref{eq:dynamical_average_MHRW}, one has to keep track of $\Delta_t E^{\text{a}}$ as well as a histogram $r_{\alpha,t}$ over the distribution of the rejected positive energy changes. For many discrete models this is a managebly small set of extra information to sample in addition to the standard observables.

\section{Testing The Method on the $2D$ Ising Model}

We test the nonequilibrium multihistogram reweighting method on a Metropolis dynamics STCD simulation of a two dimensional ferrormagnetic Ising model on a square $L \times L = 64 \times 64$ lattice with periodic boundary conditions. The Hamiltonian is given by
\begin{equation}
 H[s] = -\sum_{\avg{i,j}} s_i s_j,
\end{equation}
where $i$ and $j$ are lattice indices and $s_i \in \set{-1,1}$. Note that $\set{\Delta E} = \set{-4,-2,0,2,4}$, which means that we just need to sample three extra quantities in addition to the observable(s), in this case the dynamical average of the magnetization per site,
\begin{equation}
 m_t \equiv L^{-2}\sum_i s_{t,i}.
\end{equation}
To make the test simple, we consider only simulations from a perfectly ordered state, $\Sigma_0 = \sigma_0 = \set{s_i = 1|\forall i}$, and so we may choose $\omega(\sigma_0) = 1$.

\begin{figure}
\includegraphics{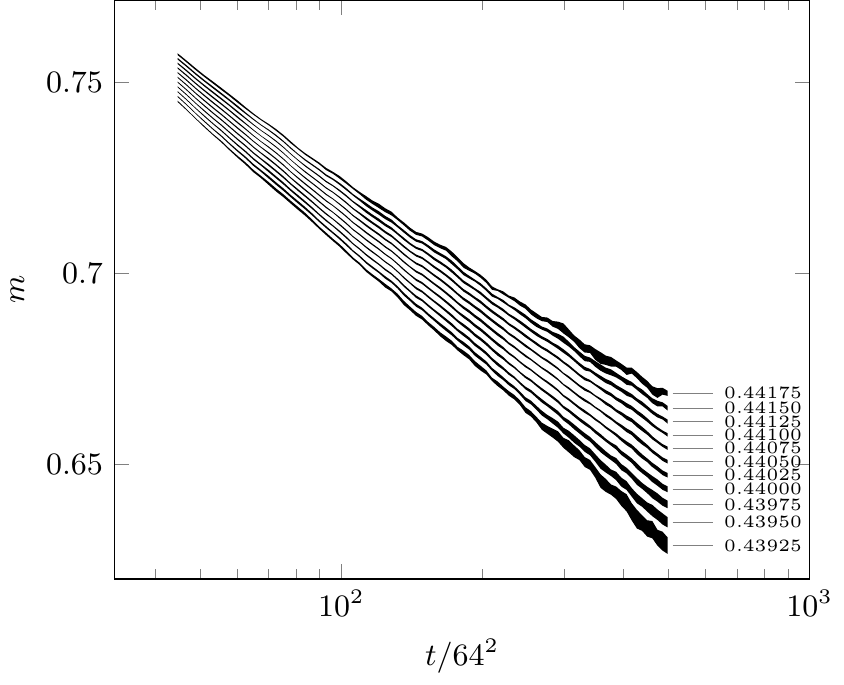}%
\caption{Plot of magnetization curves for the multihistogram reweighted couplings (from below) $\beta^* =\num{0.43925}, \num{0.43950}, \ldots, \num{0.44175}$. The reweighting is based on simulations at $\beta_q = \num{0.4400}$, $\num{0.4405}$ and $\num{0.4410}$. Line thickness corresponds to $\pm$ jackknife standard error. \label{fig:multihistogram_curves}}
\end{figure}

Figure \ref{fig:multihistogram_curves} shows a fan of reweighted magnetization curves as a function of Monte Carlo time, obtained from three simulations performed at $\beta_1 = \num{0.4400}$, $\beta_2 = \num{0.4405}$, and $\beta_3 = \num{0.4410}$. (This is close to criticality, $\beta_\text{c} = \ln(1 + \sqrt{2})/2 = \num{0.4400687}\ldots$) $N_q = \num{20000} \ \forall q$. Errors are obtained by the Jackknife method. Notice how the error grows as the reweighted coupling deviates more from the simulated couplings and the overlap in energy histograms worsens, as is the case for all statistical reweighting techniques.

\begin{figure}
\includegraphics{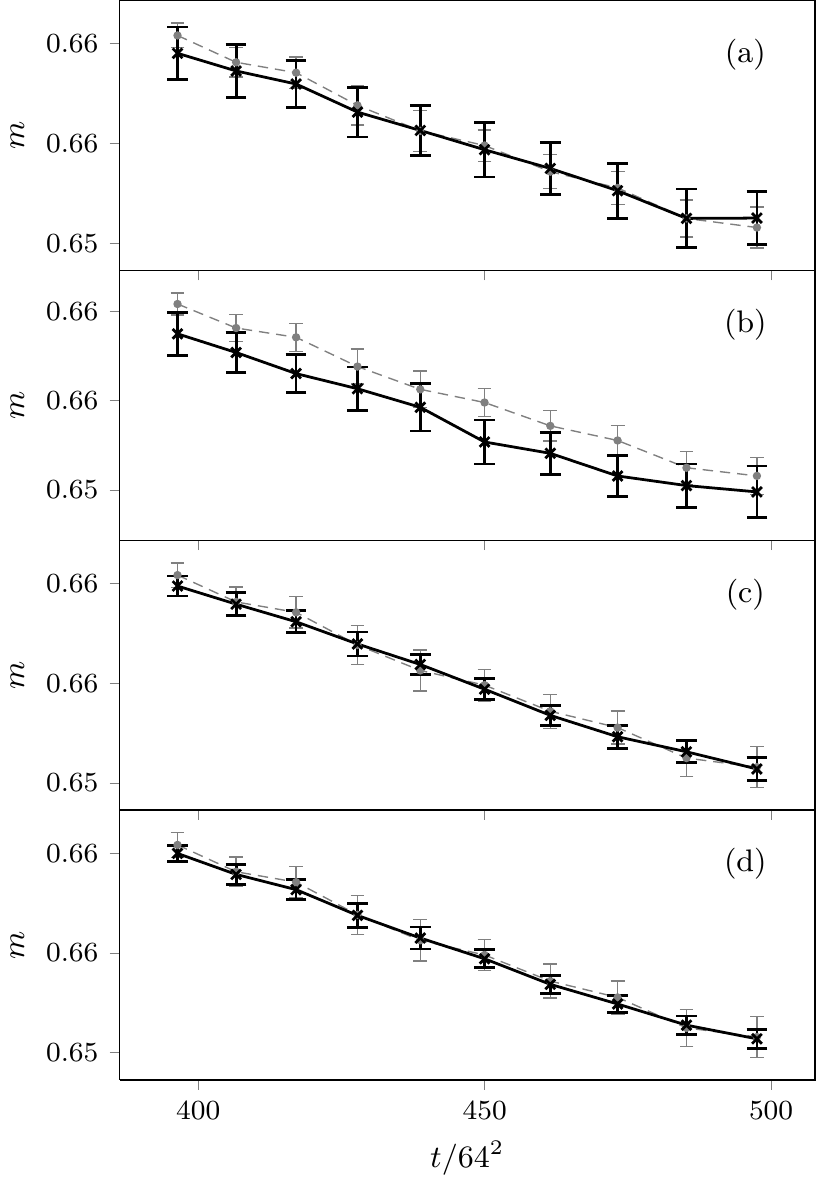}%
\caption{Comparison between non-reweighted magnetizations curves (dashed gray lines) at $\beta = \num{0.4405}$ and reweighted curves (thick black lines), reweighted to $\beta^* = \num{0.4405}$. (a) shows the single histogram reweighting from a simulation at $\beta_q = \num{0.4400}$. (b) shows the single histogram reweighting from a simulation at $\beta_q = \num{0.4410}$. (c) shows a multihistogram reweighting based on both these simulations. (d) shows a multihistogram reweighting using all data available, i.e. the simulations as $\beta_q = \num{0.4400}$, $\num{0.4405}$ and $\num{0.4410}$. Error bars are jackknife standard errors.\label{fig:technique_comparison}}
\end{figure}

We compare the multihistogram reweighting with raw data and single histogram reweighting. First we perform single histogram reweighting of the $q = 1$ simulation to $\beta^* = \beta_2$ and from simulation $3$ to $\beta^* = \beta_2$, then multihistogram reweighting to $\beta^* = \beta_2$ using simulation $1$ and $2$, and finally multihistogram reweighting to $\beta^* = \beta_2$ using the entire set of available data, $q = 1,2,3$. The results are compared with the non-reweighted dynamical average of the raw data of simulation 2 in Fig. \ref{fig:technique_comparison}. Notice that while the single histogram reweighting always performs worse (in terms larger error bars) than the raw data, the multihistogram reweighting may even outperform the raw data, given that, like in this case, more simulations with overlapping energy histograms exist.

It should be noted that the computational cost of multihistogram reweighting is of the same order of magnitude as single histogram reweighting -- and negligible compared to the simulations.

\section{Conclusion}

To summarize, we have generalized the single histogram nonequilibrium Markov Chain reweighting technique of Yin et al.~\cite{PhysRevE.72.036122} to a multihistogram framework based on the equilibrium method of Ferrenberg and Swendsen. In this manner we can take advantage of several nonequilibirum simulations performed at different (but close) couplings, producing dynamical averages with equal or smaller errors than those obtained by averages based on the non-reweighted or single histogram reweighted datasets.

\begin{acknowledgments}
The author thanks E. V. Herland for useful discussions and feedback, NTNU for financial support and the Norwegian consortium for high-performance computing (NOTUR) for (a very modest use of) computer time.
\end{acknowledgments}

\bibliography{references}

\begin{thebibliography}{6}%
\makeatletter
\providecommand \@ifxundefined [1]{%
 \@ifx{#1\undefined}
}%
\providecommand \@ifnum [1]{%
 \ifnum #1\expandafter \@firstoftwo
 \else \expandafter \@secondoftwo
 \fi
}%
\providecommand \@ifx [1]{%
 \ifx #1\expandafter \@firstoftwo
 \else \expandafter \@secondoftwo
 \fi
}%
\providecommand \natexlab [1]{#1}%
\providecommand \enquote  [1]{``#1''}%
\providecommand \bibnamefont  [1]{#1}%
\providecommand \bibfnamefont [1]{#1}%
\providecommand \citenamefont [1]{#1}%
\providecommand \href@noop [0]{\@secondoftwo}%
\providecommand \href [0]{\begingroup \@sanitize@url \@href}%
\providecommand \@href[1]{\@@startlink{#1}\@@href}%
\providecommand \@@href[1]{\endgroup#1\@@endlink}%
\providecommand \@sanitize@url [0]{\catcode `\\12\catcode `\$12\catcode
  `\&12\catcode `\#12\catcode `\^12\catcode `\_12\catcode `\%12\relax}%
\providecommand \@@startlink[1]{}%
\providecommand \@@endlink[0]{}%
\providecommand \url  [0]{\begingroup\@sanitize@url \@url }%
\providecommand \@url [1]{\endgroup\@href {#1}{\urlprefix }}%
\providecommand \urlprefix  [0]{URL }%
\providecommand \Eprint [0]{\href }%
\providecommand \doibase [0]{http://dx.doi.org/}%
\providecommand \selectlanguage [0]{\@gobble}%
\providecommand \bibinfo  [0]{\@secondoftwo}%
\providecommand \bibfield  [0]{\@secondoftwo}%
\providecommand \translation [1]{[#1]}%
\providecommand \BibitemOpen [0]{}%
\providecommand \bibitemStop [0]{}%
\providecommand \bibitemNoStop [0]{.\EOS\space}%
\providecommand \EOS [0]{\spacefactor3000\relax}%
\providecommand \BibitemShut  [1]{\csname bibitem#1\endcsname}%
\let\auto@bib@innerbib\@empty
\bibitem [{\citenamefont {Ozeki}\ and\ \citenamefont
  {Ito}(2007)}]{Ozeki_Ito_2007}%
  \BibitemOpen
  \bibfield  {author} {\bibinfo {author} {\bibfnamefont {Y.}~\bibnamefont
  {Ozeki}}\ and\ \bibinfo {author} {\bibfnamefont {N.}~\bibnamefont {Ito}},\
  }\href {http://stacks.iop.org/1751-8121/40/i=31/a=R01} {\bibfield  {journal}
  {\bibinfo  {journal} {Journal of Physics A: Mathematical and Theoretical}\
  }\textbf {\bibinfo {volume} {40}},\ \bibinfo {pages} {R149} (\bibinfo {year}
  {2007})}\BibitemShut {NoStop}%
\bibitem [{\citenamefont {Albano}\ \emph {et~al.}(2011)\citenamefont {Albano},
  \citenamefont {Bab}, \citenamefont {Baglietto}, \citenamefont {Borzi},
  \citenamefont {Grigera}, \citenamefont {Loscar}, \citenamefont {Rodriguez},
  \citenamefont {Puzzo},\ and\ \citenamefont {Saracco}}]{Albato_et_al_2011}%
  \BibitemOpen
  \bibfield  {author} {\bibinfo {author} {\bibfnamefont {E.~V.}\ \bibnamefont
  {Albano}}, \bibinfo {author} {\bibfnamefont {M.~A.}\ \bibnamefont {Bab}},
  \bibinfo {author} {\bibfnamefont {G.}~\bibnamefont {Baglietto}}, \bibinfo
  {author} {\bibfnamefont {R.~A.}\ \bibnamefont {Borzi}}, \bibinfo {author}
  {\bibfnamefont {T.~S.}\ \bibnamefont {Grigera}}, \bibinfo {author}
  {\bibfnamefont {E.~S.}\ \bibnamefont {Loscar}}, \bibinfo {author}
  {\bibfnamefont {D.~E.}\ \bibnamefont {Rodriguez}}, \bibinfo {author}
  {\bibfnamefont {M.~L.~R.}\ \bibnamefont {Puzzo}}, \ and\ \bibinfo {author}
  {\bibfnamefont {G.~P.}\ \bibnamefont {Saracco}},\ }\href
  {http://stacks.iop.org/0034-4885/74/i=2/a=026501} {\bibfield  {journal}
  {\bibinfo  {journal} {Reports on Progress in Physics}\ }\textbf {\bibinfo
  {volume} {74}},\ \bibinfo {pages} {026501} (\bibinfo {year}
  {2011})}\BibitemShut {NoStop}%
\bibitem [{\citenamefont {Ferrenberg}\ and\ \citenamefont
  {Swendsen}(1988)}]{PhysRevLett.61.2635}%
  \BibitemOpen
  \bibfield  {author} {\bibinfo {author} {\bibfnamefont {A.~M.}\ \bibnamefont
  {Ferrenberg}}\ and\ \bibinfo {author} {\bibfnamefont {R.~H.}\ \bibnamefont
  {Swendsen}},\ }\href {\doibase 10.1103/PhysRevLett.61.2635} {\bibfield
  {journal} {\bibinfo  {journal} {Phys. Rev. Lett.}\ }\textbf {\bibinfo
  {volume} {61}},\ \bibinfo {pages} {2635} (\bibinfo {year}
  {1988})}\BibitemShut {NoStop}%
\bibitem [{\citenamefont {Ferrenberg}\ and\ \citenamefont
  {Swendsen}(1989)}]{PhysRevLett.63.1195}%
  \BibitemOpen
  \bibfield  {author} {\bibinfo {author} {\bibfnamefont {A.~M.}\ \bibnamefont
  {Ferrenberg}}\ and\ \bibinfo {author} {\bibfnamefont {R.~H.}\ \bibnamefont
  {Swendsen}},\ }\href {\doibase 10.1103/PhysRevLett.63.1195} {\bibfield
  {journal} {\bibinfo  {journal} {Phys. Rev. Lett.}\ }\textbf {\bibinfo
  {volume} {63}},\ \bibinfo {pages} {1195} (\bibinfo {year}
  {1989})}\BibitemShut {NoStop}%
\bibitem [{\citenamefont {Lee}\ and\ \citenamefont
  {Okabe}(2005)}]{PhysRevE.71.015102}%
  \BibitemOpen
  \bibfield  {author} {\bibinfo {author} {\bibfnamefont {H.~K.}\ \bibnamefont
  {Lee}}\ and\ \bibinfo {author} {\bibfnamefont {Y.}~\bibnamefont {Okabe}},\
  }\href {\doibase 10.1103/PhysRevE.71.015102} {\bibfield  {journal} {\bibinfo
  {journal} {Phys. Rev. E}\ }\textbf {\bibinfo {volume} {71}},\ \bibinfo
  {pages} {015102} (\bibinfo {year} {2005})}\BibitemShut {NoStop}%
\bibitem [{\citenamefont {Yin}\ \emph {et~al.}(2005)\citenamefont {Yin},
  \citenamefont {Zheng},\ and\ \citenamefont {Trimper}}]{PhysRevE.72.036122}%
  \BibitemOpen
  \bibfield  {author} {\bibinfo {author} {\bibfnamefont {J.~Q.}\ \bibnamefont
  {Yin}}, \bibinfo {author} {\bibfnamefont {B.}~\bibnamefont {Zheng}}, \ and\
  \bibinfo {author} {\bibfnamefont {S.}~\bibnamefont {Trimper}},\ }\href
  {\doibase 10.1103/PhysRevE.72.036122} {\bibfield  {journal} {\bibinfo
  {journal} {Phys. Rev. E}\ }\textbf {\bibinfo {volume} {72}},\ \bibinfo
  {pages} {036122} (\bibinfo {year} {2005})}\BibitemShut {NoStop}%
\end{thebibliography}%

\end{document}